\documentclass[smallcondensed]{svjour3}
\usepackage{amssymb}
\usepackage{amsfonts}
\usepackage{amsmath}
\usepackage{fancyhdr}
\usepackage{epsfig}
\usepackage{graphicx}
\usepackage {longtable}
\usepackage{lineno,hyperref}
\smartqed  

\RequirePackage{fix-cm}

\newcommand{\Bc}{\ensuremath{\mathcal  B}}
\newcommand{\Cc}{\ensuremath{\mathcal  C}}

\newcommand{\Fc}{\ensuremath{\mathcal  F}}
\newcommand{\Gc}{\ensuremath{\mathcal  G}}

\newcommand{\Sc}{\ensuremath{\mathcal  S}}

\newcommand{\Vc}{\ensuremath{\mathcal  V}}

\newcommand{\ab}{\ensuremath{\mathbf{a}}}
\newcommand{\cb}{\ensuremath{\mathbf{c}}}
\newcommand{\eb}{\ensuremath{\mathbf{e}}}
\newcommand{\gb}{\ensuremath{\mathbf{g}}}
\newcommand{\hb}{\ensuremath{\mathbf{h}}}
\newcommand{\mb}{\ensuremath{\mathbf{m}}}

\newcommand{\ub}{\ensuremath{\mathbf{u}}}
\newcommand{\vb}{\ensuremath{\mathbf{v}}}
\newcommand{\wb}{\ensuremath{\mathbf{w}}}
\newcommand{\xb}{\ensuremath{\mathbf{x}}}

\newcommand{\Ab}{\ensuremath{\mathbf{A}}}
\newcommand{\Bb}{\ensuremath{\mathbf{B}}}
\newcommand{\Gb}{\ensuremath{\mathbf{G}}}

\newcommand{\Pb}{\ensuremath{\mathbf{P}}}

\newcommand{\bF}{\ensuremath{\mathbb{F}}}

\newcommand{\bP}{\ensuremath{\mathbb{P}}}

\newcommand{\beg}{\begin{example}}
\newcommand{\eeg}{\end{example}}
\newcommand{\bit}{\begin{itemize}}
\newcommand{\eit}{\end{itemize}}
\newcommand{\bcor}{\begin{corollary}}
\newcommand{\ecor}{\end{corollary}}
\newcommand{\beq}{\begin{equation}}
\newcommand{\eeq}{\end{equation}}
\newcommand{\beqn}{\begin{equation*}}
\newcommand{\eeqn}{\end{equation*}}
\newcommand{\bea}{\begin{eqnarray}}
\newcommand{\eea}{\end{eqnarray}}
\newcommand{\bean}{\begin{eqnarray*}}
\newcommand{\eean}{\end{eqnarray*}}
\newcommand{\ben}{\begin{enumerate}}
\newcommand{\een}{\end{enumerate}}
\newcommand{\bdefn}{\begin{definition}}
\newcommand{\edefn}{\end{definition}}
\newcommand{\bnote}{\begin{note}}
\newcommand{\enote}{\end{note}}
\newcommand{\bprop}{\begin{proposition}}
\newcommand{\eprop}{\end{proposition}}
\newcommand{\blem}{\begin{lemma}}
\newcommand{\elem}{\end{lemma}}
\newcommand{\bthm}{\begin{theorem}}
\newcommand{\ethm}{\end{theorem}}

\newcommand{\qrk}{\mathrm{rk}_q}

\newcommand{\bfm}{\mathbb{F}_{q^m}}

\newcommand{\cpr}{\mathbf{\Cc}_{pub}^{\perp}}

\begin{document}

\title{Extending Coggia-Couvreur Attack on Loidreau's Rank-metric Cryptosystem }




\author{ Anirban Ghatak }


\institute{Anirban Ghatak \at
	R. C. Bose Centre for Security and Cryptology \\
	ISI, Kolkata, India
	Tel.: +91 (33) 2575 2012\\
	\email{ghatak.anirban@gmail.com}           
}


\maketitle

		\begin{abstract}
			A recent paper by Coggia and Couvreur presents a polynomial time key-recovery attack on Loidreau's encryption scheme, based on rank-metric codes, for some parameters. Their attack was formulated for the particular case when the secret matrix in Loidreau's scheme is restricted to a $ 2 $-dimensional subspace. We present an extension of the Coggia-Couvreur attack to deal with secret matrices chosen over subspaces of dimension greater than $ 2 $. 
		\end{abstract}
		

\section{Introduction}
One of the directions of current research in code-based cryptography is to formulate a strong variant of the McEliece scheme \cite{mcEl} using codes in the rank-metric. The majority of proposals for rank-metric cryptosystems have been based on the use of Gabidulin codes \cite{gpt} and their variants \cite{purewah} or low-rank parity check (LRPC) codes \cite{gaborlrpc}. Of these, most of the Gabidulin-code based cryptosystems have been subjected to successful key-recovery attacks, for instance, R. Overbeck's attack \cite{overbeck} on the Gabidulin-Paramonov-Tretjakov (GPT) proposal (\cite{gpt}). The basis of Overbeck's attack is the fact that the application of a Frobenius-type map on a Gabidulin code generator matrix can be used to distinguish it from a random matrix. This principle - referred to in literature as a ``distinguisher'' - has since been used repeatedly to mount successful key recovery attacks on repair proposals on the GPT and other rank-metric variants; for instance, the attack on Faure-Loidreau's scheme \cite{fauloi} by Gaborit \emph{et al.} \cite{gok}.\\
It follows that the first design objective of any rank-metric cryptosystem, based on Gabidulin-type codes, is resistance to key recovery attack along the lines of Overbeck's method. So far, a few recent proposals claim to have achieved that - for example, the repair of the Faure-Loidreau rank-metric scheme by Wachter-Zeh \emph{et al.} \cite{flrep} and Loidreau's scheme \cite{loi}. Loidreau's scheme uses Gabidulin codes for encryption with the following additional feature. It uses a secret matrix with entries from a strict subspace of the field underlying the Gabidulin code. It is claimed that Overbeck's Frobenius-map distinguisher fails if the Gabidulin generator matrix is modified with this secret matrix. Coggia and Couvreur have shown (\cite{coco}) that polynomial time key recovery is possible with Loidreau's scheme, for certain parameter constraints, when the dimension of the secret subspace is precisely $ \lambda = 2 $. While the dimension constraint appears restrictive, their approach has opened up the possibility for cryptanalysis of rank-metric schemes which have claimed resistance to attacks using Overbeck-type distinguishers. In this article we extend the Coggia-Couvreur key-recovery attack on Loidreau's cryptosystem to admit secret matrices over subspaces of dimension $ \lambda = 3 $.\\

\textbf{Contributions:}
\begin{enumerate}
	\item A proof of the non-random nature of the public generator matrix (i.e. formulating a ``distinguisher'' as in \cite{coco}) in Loidreau's scheme for $ \lambda \geq 3 $.
	\item Completing the key-recovery attack for $ \lambda =3 $.
\end{enumerate}

\textit{Organization of the article:} The first section outlines Loidreau's scheme and describes the steps of the Coggia-Couvreur attack. Section \ref{dist} formulates the distinguisher for any dimension of the secret subspace. The next section (Section \ref{secsubsp}) deals with the computation of certain specific subspaces, which are subsequently used in the extraction of parameters. Finally Section \ref{secattaq} provides the details of extending the key recovery attack to the case of $ \lambda =3 $. We conclude with a discussion on the results and future work. 
\section{Loidreau's Scheme and Coggia-Couvreur Attack}\label{loco}
We first outline Loidreau's scheme and discuss the reason it is claimed to resist Overbeck's distinguisher.
\subsection{Outline of Loidreau's scheme:}
Loidreau's scheme (\cite{loi}) is similar to the Gabidulin rank-metric scheme, modified to resist Overbeck's distinguisher. 
\bit
\item $ \Gb $ generator matrix of a Gabidulin code $ \Gc_k(\ab) $ over $ \bfm $; $ \qrk (\ab) = n $.
\item $ \Vc \subset \bfm$, $ \dim_q (\Vc) = \lambda \leq m $; $ \Pb \in GL_n (q^m) $ is a matrix over $\Vc$. 
\item  Define $ \Gb_p := \Gb \Pb ^{-1} $ and $ t := \lfloor \frac{n-k}{2\lambda}\rfloor $.
\item The public key is $ K_p := (\Gb_p, t) $ and the secret key, $ K_s := (\ab, \Pb)  $.
\item \textbf{Encryption:} $ \cb = \mb \Gb_p + \eb $; $ \eb \in \bfm^{n}  $ and $ \qrk (\eb) = t $.
\item \textbf{Decryption:} $ \cb \Pb = \mb \Gb + \eb \Pb $.\\
The $ \bF_q $-dimension of the product space $ \mathrm{supp}(\eb). \Vc $ is $ t\lambda \leq \lfloor \frac{n-k}{2}\rfloor $, and hence, decoding for $ \Gc_k(\ab) $ will extract plaintext $ \mb $. 
\eit
Overbeck's attack relies on the ``Frobenius map'' distinguisher on a Gabidulin code structure of the public generator matrix. Raising elements of $ \Gb_p $ to successive $ q $-powers and vertically stacking the rows results in an augmented matrix where the increase of rank is only by unity at each stage. But the rank of such a matrix constructed from a random code matrix would have an increment equal to the rank of the code matrix at every stage with high probability. So the Gabidulin type code matrix will have markedly less rank at some stage of the augmentation. However, in Loidreau's scheme, $ \Gb_p = \Gb\Pb^{-1} $, where $ \Pb $ is constrained to some subspace $ \Vc \subset \bfm $. But there is no control over the entries of $ \Pb ^{-1} $ and so, the rank increment achieved at each stage via the $ q $-exponentiation map is no longer exactly unity. Hence Overbeck's distinguisher is no longer effective.
\subsection{Coggia-Couvreur Attack for $ \lambda =2 $}
Coggia and Couvreur defined a distinguisher for Loidreau's scheme when the secret subspace $ \Vc $ has dimension $ \lambda = 2 $. In this particular case they showed that key recovery is possible by solving for a triple $ (\gamma, \gb, \hb) $ over $ \bfm $, where $ \gamma $ specifies the secret subspace and $ \gb, \hb $ specify a decomposition of the (dual) public generator matrix $ \mathbf{\Cc}_{pub}^{\perp} $ in terms of the secret matrix $ \Pb $ and the secret generator matrix.\\
\emph{Notation:} Henceforth in the article, the notation $ \gb^{[i]} $, $ i $ an integer, would mean raising all components of $ \gb $ to the $ q^i $-th power.

\subsubsection{Distinguisher for $ \lambda = 2 $}

\begin{itemize}
\item Without loss of generality can specify the secret subspace as:\[ \Vc = \langle (1, \gamma) \rangle ;  \gamma \in \bfm \setminus \bF_q \] 
Then we have the formulation: $ \Pb ^T = \Pb _0 + \gamma \Pb _1$, where $ \Pb _0 , \Pb _1 \in GL_n (\bF_q) $.
\item The dual of the secret code has generator matrix: $ \cpr = \Gc_{n-k} (\ab') $ for some $ \ab' \in \bfm^n $ with $ \qrk (\ab') = n $.
\item Define $ \gb := \ab' \Pb _0  $ and  $ \hb := \ab' \Pb _1 $.
\item $ \cpr = \langle \gb + \gamma \hb, \gb^{[1]} + \gamma \hb^{[1]}, \cdots, \gb^{[r]} + \gamma \hb^{[r]} \rangle $, where $ r:= n-k-1 $.
\bthm[\cite{coco}]
\[\dim_{q^m}(\cpr + {\cpr}^{[1]} +  {\cpr}^{[2]}) \leq 2 \dim_{q^m} \cpr + 2. \]
\ethm
\end{itemize}
It follows that $ \cpr $ will be distinguishable from a random matrix in polynomial time whenever $ 2(n-k) + 2 < n $, i.e. when $ n < 2k - 2 $. Hence, for the distinguisher to work, the rate of the code should satisfy: $ k/n > 1/2 + 1/n \approx 1/2 $.

\subsubsection{Recovery of alternate key}
\begin{itemize}
	\item The following iterated intersection is shown to be of $ \bfm $-dimension $ 2 $ and is generated by $ \gb^{[r]} + \gamma^{[r]} \hb^{[r]} $ and $ \gb^{[r+1]} + \gamma^{[1]} \hb^{[r+1]} $:
	\[ (\cpr + {\cpr}^{[1]})\cap ({\cpr}^{[1]} +  {\cpr}^{[2]}) \cap \cdots \cap ({\cpr}^{[r]} +  {\cpr}^{[r+1]})\]
	\item Can extract the subspaces: $ \langle \gb + \gamma \hb \rangle $, $ \langle \gb + \gamma^{[-1]} \hb \rangle $, ..., $ \langle \gb + \gamma^{[-r]} \hb \rangle $.
	\blem[\cite{coco}]
	For $ i,j \in \{1, \cdots, r\}, i \neq j $, there exists a unique pair 
	\beqn (\ub_ {ij}, \vb_{ij}) \in \langle \gb + \gamma^{[-i]} \hb \rangle \times \langle \gb + \gamma^{[-j]} \hb \rangle \eeqn 
	such that $ \ub_ {ij} + \vb_{ij} = \gb + \gamma \hb $.
	\elem
	\item The previous lemma leads to expressing $ \ub_{ij}, \vb_{ij} $ in terms of $ \gamma, \gb, \hb $ as follows:
	\[ \ub_{ij} = \frac{\gamma^{[-j]} - \gamma}{\gamma^{[-j]} - \gamma^{[-i]}}(\gb + \gamma^{[-i]} \hb); \, \, \, \vb_{ij} = \frac{\gamma^{[-i]} - \gamma}{\gamma^{[-i]} - \gamma^{[-j]}}(\gb + \gamma^{[-j]} \hb). \]
	\item Hence for some $ \alpha_{j_1 j_2} \in \bfm $, $ \alpha_{j_1 j_2} $ a function of $ \gamma $, we get: $ \ub_{ij_1} = \alpha_{j_1 j_2} \ub_{ij_2} $, as both are $ \bfm $-multiples of $ \gb + \gamma^{[-i]} \hb $.
	\item Having computed such an $ \alpha $ for a specific pair of $ \ub_{ij}  $'s, viz. $ \ub_{12}, \ub_{13}   $, one obtains a polynomial $ P_{\gamma} (X) $, with $ \gamma $ as a root, of the following form.\\
	$ P_{\gamma} (X) = \frac{Q_{\gamma} (X)}{(X^q - X)^{q+1}} $, where 
	\[Q_{\gamma} (X) = (X^{q^3} - X^q) (X^{q^2} - X) - \alpha^{q^3}(X^{q^3} - X) (X^{q^2} - X^q) \].
	\item A crucial result (Proposition 5. \cite{coco}) shows that, viewed as vectors over $ \bF_q $, the set of roots of $ P_{\gamma} (X) $ form an orbit under the action of the projective linear group $ PGL(2,q) $. The action is sharply transitive and as such, \emph{any root} of $ P_{\gamma} (X) $ can be chosen as a valid $ \gamma $.
	\item With a valid choice for $ \gamma $, say $ \gamma ' $, and using the known quantities $ \gb + \gamma \hb $ and say, $ \ub_{12} $, set up the equations:
	\[ \gb + \gamma \hb  =  \gb' + \gamma' \hb' ;\,\,\, \ub_{12} = \frac{{\gamma'}^{[-2]} - \gamma'}{{\gamma'}^{[-2]} - {\gamma'}^{[-1]}}(\gb' + {\gamma'} ^{[-1]} \hb').\]
	Solving for the triple $ (\gamma', \gb ', \hb ') $ provides an alternative secret key.
	
\end{itemize}
\section{Distinguisher for Loidreau's Scheme for $ \lambda \geq 3 $} \label{dist}

We attempt to extend Coggia-Couvreur attack to the cases where the secret subspace $ \Vc $ has dimension $ \lambda >2 $. The first step is to establish the non-randomness of the public matrix, i.e. formulating the so-called ``distinguisher''.
\subsection{Distinguisher for $ \lambda = 3 $.}\label{subsecdefgi}
Assume that $ \Vc = \langle 1, \gamma_1, \gamma_2 \rangle $, where $ \gamma_i \in \bfm \setminus \bF_q $.\\
Accordingly we have: $ \Pb ^T = \Pb _0 + \gamma_1 \Pb _1 + \gamma_2 \Pb _2 $; where all $ \Pb _i \in GL_n (q) $.\\
Define $ \gb _0 = \ab \Pb_0, \,  \gb _1 = \ab \Pb_1,  \, \gb _2 = \ab \Pb_2$. Then we have, for $ r:= n-k-1 $,
 \[ \cpr = \langle (\gb_0 + \gamma_1 \gb_1 + \gamma_2 \gb_2) ,  (\gb_{0}^{[1]} + \gamma_1 \gb_{1}^{[1]} + \gamma_2 \gb_{2}^{[1]} ) , \cdots, (\gb_{0}^{[r]} + \gamma_1 \gb_{1}^{[r]} + \gamma_2\gb_{2}^{[r]}) \rangle \]
 Likewise $ {\cpr}^{[1]} $ is spanned by:
 \[(\gb_{0}^{[1]} + \gamma_{1}^{[1]} \gb_{1}^{[1]} + \gamma_{2}^{[1]} \gb_{2}^{[1]}) ,  (\gb_{0}^{[2]} + \gamma_{1}^{[1]} \gb_{1}^{[2]} + \gamma_{2}^{[1]} \gb_{2}^{[2]} ) , \cdots, (\gb_{0}^{[r+1]} + \gamma_{1}^{[1]} \gb_{1}^{[r+1]} + \gamma_{2}^{[1]}\gb_{2}^{[r+1]})\]
 and $ {\cpr}^{[2]} $ is spanned by:
 \[(\gb_{0}^{[2]} + \gamma_{1}^{[2]} \gb_{1}^{[2]} + \gamma_{2}^{[2]} \gb_{2}^{[2]}) ,  (\gb_{0}^{[3]} + \gamma_{1}^{[2]} \gb_{1}^{[3]} + \gamma_{2}^{[2]} \gb_{2}^{[3]} ) , \cdots, (\gb_{0}^{[r+2]} + \gamma_{1}^{[2]} \gb_{1}^{[r+2]} + \gamma_{2}^{[2]}\gb_{2}^{[r+2]}).\]
 
 Hence, akin to the formulation for $ \lambda =2 $ presented in \cite{coco}, we state the following theorem based on the preceding discussion\footnote{In a recent version of their paper: arxiv.org/abs/1903.02933v2, the authors have indicated the form of the sum space to extend their argument for $ \lambda =2 $. We had independently arrived at a similar conclusion based on the original version of their paper and have, moreover, presented the details of the proof for $ \lambda \geq 3 $.}.
 
 \bthm \label{3int}
 The dual $ \cpr$ of the public code in Loidreau's scheme satisfies: 
 \[\dim_{q^m} (\cpr + {\cpr}^{[1]} + {\cpr}^{[2]} + {\cpr}^{[3]} ) \leq  3\dim_{q^m} \cpr + 3. \]
 \ethm
  
 \begin{proof}
 Consider the sum space $ \cpr + {\cpr}^{[1]} + {\cpr}^{[2]} $. For $ i = 2, \cdots, r  $, given the choice of $ \gamma_1, \gamma_2 \in \bfm \setminus \bF_q$, the following matrix is invertible.
 \[ \begin{pmatrix}
 1 & \gamma_1 & \gamma_2\\
 1 & \gamma_{1}^{[1]} & \gamma_{2}^{[1]}\\
 1 & \gamma_{1}^{[2]} & \gamma_{2}^{[2]}
 \end{pmatrix}\] 
 It follows that one can extract the triples $ (\gb_{0}^{[i]}, \gb_{1}^{[i]}, \gb_{2}^{[i]} ) $, $ i = 2, \cdots, r  $, from \\
 
 $ (\gb_{0}^{[i]} + \gamma_1 \gb_{1}^{[i]} + \gamma_2\gb_{2}^{[i]}) $, $ (\gb_{0}^{[i]} + \gamma_{1}^{[1]} \gb_{1}^{[i]} + \gamma_{2}^{[1]}\gb_{2}^{[i]}) $  and $ (\gb_{0}^{[i]} + \gamma_{1}^{[2]} \gb_{1}^{[i]} + \gamma_{2}^{[2]}\gb_{2}^{[i]}) $.\\
  
 In addition to these $ (n-k-2) $ triples, $ \cpr + {\cpr}^{[1]} + {\cpr}^{[2]} $ contains the following $ 6 $ vectors:\\
 
 $ (\gb_0 + \gamma_1 \gb_1 + \gamma_2 \gb_2) ,  \, \, (\gb_{0}^{[1]} + \gamma_1 \gb_{1}^{[1]} + \gamma_2 \gb_{2}^{[1]} ),\,\,  (\gb_{0}^{[1]} + \gamma_{1}^{[1]} \gb_{1}^{[1]} + \gamma_{2}^{[1]} \gb_{2}^{[1]}) $\\
 
 $ (\gb_{0}^{[r+1]} + \gamma_{1}^{[1]} \gb_{1}^{[r+1]} + \gamma_{2}^{[1]}\gb_{2}^{[r+1]}), \, (\gb_{0}^{[r+1]} + \gamma_{1}^{[2]} \gb_{1}^{[r+1]} + \gamma_{2}^{[2]}\gb_{2}^{[r+1]}), $ \\
 
 \indent and $ (\gb_{0}^{[r+2]} + \gamma_{1}^{[2]} \gb_{1}^{[r+2]} + \gamma_{2}^{[2]}\gb_{2}^{[r+2]}). $\\
  
 Thus we can conclude that the sum space $ \cpr + {\cpr}^{[1]} + {\cpr}^{[2]} $ is spanned by $ 3(n-k-2) + 3 + 3 = 3(n-k)$ vectors as outlined above. Adding $ {\cpr}^{[3]} $ to this sum space adds vectors involving terms of $ q  $-power $ 3 $ and above, going up to the term having $ (r + 3) $-th power of $ q $.\\
 Evidently this allows the extraction of another triple  $ \{ \gb_{0}^{[r+1]} , \gb_{1}^{[r+1]}, \gb_{2}^{[r+1]} \} $, adds two terms with $ q $-power $ r + 2 $ and a last term with $ q $-power $ r+ 3 $.\\ 
 Therefore,
  \[\dim_{q^m} (\cpr + {\cpr}^{[1]} + {\cpr}^{[2]} + {\cpr}^{[3]} ) \leq  3(n-k) + 3. \]

\end{proof}
\qed

 From the above theorem, we can infer that for $ \lambda =3 $, $ \cpr $ is distinguishable in polynomial time from a random code matrix when $ 3(n-k) +3 < n$, i.e. when $ 3k -3 > 2n $. This also implies that this distinguisher is effective if the underlying codes have rate $ \frac{k}{n} > \frac{2}{3} $.
 \subsection{Distinguisher for $ \lambda>3 $}
 Based on the principle outlined for $ \lambda\ \leq 3 $, one can propose distinguishers for Loidreau's scheme, subject to a constraint on the rate $ \frac{k}{n} $ of the underlying code.
 For $ \lambda = m >3 $, it is assumed that the secret subspace $ \Vc = \langle 1, \gamma_1, \cdots, \gamma_{m-1} \rangle $, $ \gamma_i \in \bfm \setminus \bF_q $. In the same spirit as before, we first look at the $ m $-fold $ q $-power sum of $ \cpr $, the dual public code, given by: 
 \beq \label{msum} 
 \cpr + {\cpr}^{[1]} + {\cpr}^{[2]} + \cdots + {\cpr}^{[m-1]} .
 \eeq 
 Grouping together terms involving the same $ q $-powers, we can extract $ m $-tuples $ (\gb_{0}^{[i]}, \gb_{1}^{[i]},  \cdots, \gb_{m-1}^{[i]}) $ whenever we can form an invertible $ m \times m$ matrix of the following form:  
 \[\begin{pmatrix}
 1 & \gamma_1 & \cdots & \gamma_{m-1}\\
  1 & \gamma_{1}^{[1]} & \cdots & \gamma_{m-1}^{[1]}\\
  \vdots & \vdots & \ddots & \vdots\\
  1 & \gamma_{1}^{[m-1]} & \cdots & \gamma_{m-1}^{[m-1]}
 \end{pmatrix}\]
 Thus, we can count the number of terms in the $ m $-fold sum, prior to the stage that such an invertible matrix can be formed to extract the first set of m-tuples with largest $ q $-power $ m-1 $ as follows: \\
 There are precisely $  \frac{m(m-1)}{2}$ terms of the form $ (\gb_{0}^{[i]} + \gamma_{1}^{[j]} \gb_{1}^{[i]} + \cdots + \gamma_{m-1}^{[j]}\gb_{m-1}^{[i]}) $, with both $ i,j $ allowed to assume appropriate values between $ 0 $ and $ m-2 $.\\
  
Next, assuming $ m < n-k-1 $, we can continue to collect $ m $-tuples of higher $ q $-powers till $ i = n-k-1 $. Beyond this, we revert back to the sum vectors involving higher $ q $-powers all the way up to $ i = n-k-1 + (m -1) = n-k-m-2 $ and this adds another set of $ (m-1) + (m-2) + \cdots + 1 =  \frac{m(m-1)}{2}$ vectors.\\
 Therefore, we conclude that the $ m $-fold sum in (\ref{msum}) has $ \bfm $-dimension $ M $, where  
 \[ M \leq 2 \times \frac{m(m-1)}{2} + m(n-k-1 - (m-2)) = m(n-k) = m \dim_{q^m} (\cpr).\]
Adding $ {\cpr}^{[m]} $ to the sum space in (\ref{msum}) does not change the stage at which the first $ m $-tuple can be extracted. But it does alter the stage of extracting the final $ m $-tuple: we can obtain an invertible $ m \times m $ matrix to extract the tuple $ (\gb_{0}^{[n-k]}, \gb_{1}^{[n-k]},  \cdots, \gb_{m-1}^{[n-k]}) $. Beyond this, there are again a set of $ \frac{m(m-1)}{2} $ vectors with terms of increasing $ q $-powers till $ n-k-1 + m  $.  This yields a total of $ m(m-1) + m(n-k - (m-2)) = m(n -k) + m $ vectors.\\
Hence we have the following 
\bthm 
If the secret subspace of Loidreau's scheme has dimension given by $ \lambda = m \geq 2 $, the dual of the public code, denoted $ \cpr$, satisfies: 
\[\dim_{q^m} (\cpr + {\cpr}^{[1]} + {\cpr}^{[2]} + \cdots + {\cpr}^{[m]} ) \leq  m\dim_{q^m} \cpr + m .\]
\ethm
Evidently this procedure yields a distinguisher when $ m(n-k) + m <n $. So the distinguisher is effective when the underlying code has rate $ \frac{k}{n} > \frac{m-1}{m} $. This argues in favour of using low or moderate rate codes in conjunction with secret subspaces of large dimensions in order to counter this distinguisher.

\section{Computing the extraction subspaces}\label{secsubsp}
To extend the Coggia-Couvreur attack to the case $ \lambda =3 $, we attempt to obtain an alternative tuple $ \{\gamma'_{1}, \gamma'_{2}, \gb'_{0}, \gb'_{1}, \gb'_{2} \} $, which can lead to a valid secret key. Following the procedure outlined in Section \ref{loco} , the first step is to compute the subspace $ \langle \gb_0,  \gb_1, \gb_2 \rangle $, and hence, sum spaces of the form $ \langle \gb_{0} + \gamma_{1}^{[-i]} \gb_{1} + \gamma_{2}^{[-i]} \gb_{2} \rangle $, for integers $ i >0 $. These subspaces, taken together, are then utilized to extract alternative tuples for a valid secret key - hence we term them \emph{extraction subspaces}. Further we term the subspaces formed by adjoining successive $ q $-powers of subspaces as \emph{sumspaces}.
To obtain the extraction subspaces, we first examine the intersections for the $ 3 $-fold sumspaces.
\subsection{Intersections of sumspaces}
We now establish the dimensions of intersection spaces among the sumspaces with different sequences of $ q $-powers of the dual code $ \cpr $. For that, we introduce the following notation to denote sumspaces involving several $ q $-powers. Define
\[ \Sc_{j}^{i} :=  {\cpr}^{[j]} + {\cpr}^{[j+1]} + \cdots + {\cpr}^{[j+i-1]}  \]
which starts with $ q^j $-th power of $ \cpr $ and adds $ i-1 $ more terms with increasing $ q $-powers till  $ {\cpr}^{[j+i-1]} $. In this notation, we have:
\[ \Sc_{0}^{4} := (\cpr + {\cpr}^{[1]} + {\cpr}^{[2]} + {\cpr}^{[3]} ). \]
Hence, we have established in Theorem \ref{3int} that: \[
\dim_{q^m} (\Sc_{0}^{4} ) \leq  3\dim_{q^m} \cpr + 3.\]
 
Looking at the spanning sets for the $ 3 $-fold sumspaces  $ \Sc_{0}^{3} $ and $ \Sc_{1}^{3} $, it is evident that both of them have $ \bfm $-dimensions $ \leq 3(n-k) $. Assuming both the above sumspaces possess maximum dimension and further, the $ 4 $-fold sumspace $  \Sc_{0}^{4}  $ has dimension $ 3(n-k) + 3 $, we have:
	\beq\label{1st3}
	\dim_{q^m}(\Sc_{0}^{3}  \cap \Sc_{1}^{3}) = 3(n-k) - 3
  \eeq
A spanning set for $ \Sc_{2}^{3} $ consists of triples $ \{ g_{0}^{[i]}, g_{1}^{[i]}, g_{2}^{[i]}\} $ for $ i =4, 5, \cdots, r+2  $, along with the following $ 6 $ vectors:\\
\vspace{1 mm}

\noindent
$ (\gb_{0}^{[2]} + \gamma_{1}^{[2]} \gb_{1}^{[2]} +\gamma_{1}^{[2]} \gb_{2}^{[2]} ), (\gb_{0}^{[3]} + \gamma_{1}^{[2]} \gb_{1}^{[3]} + \gamma_{2}^{[2]} \gb_{2}^{[3]}), (\gb_{0}^{[3]} + \gamma_{1}^{[3]} \gb_{1}^{[3]} + \gamma_{2}^{[3]} \gb_{2}^{[3]}) $\\
$ (\gb_{0}^{[r+3]} + \gamma_{1}^{[3]} \gb_{1}^{[r+3]} +\gamma_{1}^{[3]} \gb_{2}^{[r+3]} ), (\gb_{0}^{[r+3]} + \gamma_{1}^{[4]} \gb_{1}^{[r+3]} + \gamma_{2}^{[4]} \gb_{2}^{[r+3]})$ and finally,\\ $ (\gb_{0}^{[r+4]} + \gamma_{1}^{[4]} \gb_{1}^{[r+4]} + \gamma_{2}^{[4]} \gb_{2}^{[r+4]}) $.\\
\vspace{1 mm}

\noindent

In all, we have $ 3 (r +2 -3) = 3(n-k-2)$ vectors from the triples and the $ 6 $ vectors apart from them, spanning $ \Sc_{2}^{3} $. Moreover, we can list the vectors ``shared'' between $ \Sc_{2}^{3} $ and $ \Sc_{0}^{3} $ as follows.
\begin{enumerate}
	\item All the triples for indices $ i =4, \cdots, r $.
	\item The last $ 3 $ vectors of $ \Sc_{0}^{3} $ belong to the span of the last two triples of $ \Sc_{2}^{3} $.
	\item The first $  3 $ vectors of $ \Sc_{2}^{3} $ belong to the span of the first two triples of $ \Sc_{0}^{3} $. 
	
\end{enumerate}   
Therefore, the intersection space of the above $ 3 $-fold sumspaces has dimension $ 3 (r -3) + 6 $. Thus we have:
\beq\label{2nd3}
\dim_{q^m}( \Sc_{0}^{3}  \cap  \Sc_{2}^{3} ) = 3(n-k) - 6.
\eeq
\bthm\label{th:rint}
The dimension of the intersection space $ \Sc_{0}^{3}  \cap  \Sc_{m}^{3} $ over $ \bfm $ is precisely  $ 3(n-k) - 3m $ .
\ethm
\begin{proof}
	We prove the theorem by induction on $ m  $, the first $ q $-power term of the second sumspace. By the preceding discussion, the above holds for $ m = 1,2 $. Assuming it holds upto $ m -1 $, we have 
	\beqn
	\dim_{q^m}(\Sc_{0}^{3}  \cap  \Sc_{m-1}^{3}) = 3(n-k) - 3(m-1).
	\eeqn 
	Raising the first exponent to $ m $ from $m-1$ reduces one shared triple from the intersection space. However, the last three vectors of the first sumspace and the first three vectors of the second sumspace are still shared. Thus, there is a reduction of the dimension of the intersection space by precisely $ 3 $ in going from $ m-1 $ to $ m $. Hence
	\beqn
	\dim_{q^m}(\Sc_{0}^{3}  \cap  \Sc_{m}^{3}) = 3(n-k) - 3m.
	\eeqn
	
\end{proof}
\qed

\begin{corollary}\label{rint}
	Given $ r = n-k-1 $, we have:
	\beq
	\dim_{q^m}(\Sc_{0}^{3}  \cap  \Sc_{r}^{3}) = 3(n-k) - 3r = 3.
	\eeq
\end{corollary}


\subsection{Extraction subspaces from intersection between $ \cpr $  and sumspaces}
Building on the previous analysis, we now compute the extraction subspaces from the intersection of $ \cpr $ with recursively obtained subspaces.\\
From Corollary \ref{rint}, we expect to identify $ 3 $ independent vectors which span the intersection space $ \Sc_{0}^{3}  \cap  \Sc_{r}^{3} $. Two obvious choices are the vectors:
\[ \vb_1 = \gb_{0}^{[r]} + \gamma_{1}^{[r]} \gb_{1}^{[r]} +\gamma_{2}^{[r]} \gb_{2}^{[r]}, \,\,\, \vb_2 = \gb_{0}^{[r+2]} + \gamma_{1}^{[2]} \gb_{1}^{[r+2]} +\gamma_{2}^{[2]} \gb_{2}^{[r+2]}.\]

The choice of a third vector spanning $ \Sc_{0}^{3}  \cap  \Sc_{r}^{3} $, linearly independent with respect to the two above, must be from the intersection of the following subspaces:
\[\langle \gb_{0}^{[r+1]} + \gamma_{1}^{[1]} \gb_{1}^{[r+1]} +\gamma_{2}^{[1]} \gb_{2}^{[r+1]}, \, \gb_{0}^{[r+1]} + \gamma_{1}^{[2]} \gb_{1}^{[r+1]} +\gamma_{2}^{[2]} \gb_{2}^{[r+1]} \rangle \] and
\[ \langle \gb_{0}^{[r+1]} + \gamma_{1}^{[r]} \gb_{1}^{[r+1]} +\gamma_{2}^{[r]} \gb_{2}^{[r+1]}, \,\,\, \gb_{0}^{[r+1]} + \gamma_{1}^{[r+1]} \gb_{1}^{[r+1]} +\gamma_{2}^{[r+1]} \gb_{2}^{[r+1]} \rangle. \]

Thus the third vector can have the following equivalent representations:
\beqn
\begin{split}
	\vb_3 & = \gb_{0}^{[r+1]} + (k_1 \gamma_{1}^{[1]} + k_2 \gamma_{1}^{[2]}) \gb_{1}^{[r+1]} + (k_1 \gamma_{2}^{[1]}  + k_2 \gamma_{2}^{[2]}) \gb_{2}^{[r+1]} \\
	& = \gb_{0}^{[r+1]} + (m_1 \gamma_{1}^{[r]} + m_2 \gamma_{1}^{[r +1]}) \gb_{1}^{[r+1]} + (m_1 \gamma_{2}^{[r]}  + m_2 \gamma_{2}^{[r+1]}) \gb_{2}^{[r+1]}
\end{split}
\eeqn
where $ k_1, k_ 2 , m_1, m_2 \in \bfm $. Let 
\beqn
\begin{aligned}
	\xb_1 := \vb_1 ^{[-r]} =  \gb_{0} + \gamma_{1}\gb_{1} + \gamma_{2} \gb_{2}; \\ 
	\xb_2 := \vb_2 ^{[-r]} = \gb_{0}^{[2]} + \gamma_{1}^{[2-r]} \gb_{1}^{[2]} +\gamma_{2}^{[2-r]} \gb_{2}^{[2]};  \\ 
	\xb_3 := \vb_3 ^{[-r]} = \gb_{0}^{[1]} + (a_1 \gamma_{1} + a_2 \gamma_{1}^{[1]}) \gb_{1}^{[1]} + (a_1 \gamma_{2}  + a_2 \gamma_{2}^{[1]}) \gb_{2}^{[1]}. 
\end{aligned}
\eeqn

Define $ \Bc_1 := \langle \xb_1, \xb_2, \xb_3 \rangle .$ In a manner similar to that outlined in \cite{coco}, we proceed to first obtain the subspace $ \langle \gb_0, \gb_1, \gb_2 \rangle $ and then the other extraction subspaces.
\begin{itemize}
	\item Obtain $ \langle \gb_{0} + \gamma_{1}\gb_{1} + \gamma_{2} \gb_{2} \rangle  $ from the intersection $ \cpr \cap \Bc_1 $.\\
	 Raising to the $ q$-th power, we get $ \langle \gb_{0}^{[1]} + \gamma_{1}^{[1]} \gb_{1}^{[1]} +\gamma_{2}^{[1]} \gb_{2}^{[1]} \rangle $.
	\item Consider the following sum of subspaces:
	\begin{multline*}
		 \Bc_2 = \Bc_1 + \langle \gb_{0}^{[1]} + \gamma_{1}^{[1]} \gb_{1}^{[1]} +\gamma_{2}^{[1]} \gb_{2}^{[1]} \rangle + \\ \langle \gb_{0}^{[1]}   + 
		   (b_1 \gamma_{1}^{[1-r]} + b_2 \gamma_{1}^{[2-r]}) \gb_{1}^{[1]} + (a_1 \gamma_{1}^{[1-r]}  + b_2 \gamma_{2}^{[2-r]}) \gb_{2}^{[1]}\rangle
    \end{multline*}
     where one of the forms of $\vb_3$ yields the third component.\\
     Evidently $ \Bc_2 = \langle \xb_1 , \gb_{0}^{[1]},  \gb_{1}^{[1]}, \gb_{2}^{[1]}, \xb_2 \rangle $.
     \item We can thus extract $ \langle \gb_{0}^{[1]} + \gamma_{1}^{[-1]} \gb_{1}^{[1]} +\gamma_{2}^{[-1]} \gb_{2}^{[1]} \rangle $ from $ \Bc_2 \cap (\cpr) ^{[-1]} $. Thence we obtain $ \Bc_3 := \langle \gb_{0} + \gamma_{1}^{[-2]} \gb_{1} +\gamma_{2}^{[-2]} \gb_{2} \rangle $.
     \item The following sum of subspaces:
     \[ \Bc_3 + \langle \xb_1 \rangle + \langle \gb_{0} + ( c_1 \gamma_{1}^{[-1]} + c_2 \gamma_{1}) \gb_{1} + (c_1 \gamma_{2}^{[-1]} + c_2 \gamma_{2}) \gb_{2} \rangle \]
      where the third component is obtained from $ \xb_3 $, yields $ \Bc = \langle \gb_0, \gb_1, \gb_2 \rangle $. The $ i $-th extraction subspace $ \langle \gb_{0} + \gamma_{1}^{[-i]} \gb_{1} +\gamma_{2}^{[-i]} \gb_{2} \rangle $ can be obtained by taking the $ q^{-i} $-th power of the intersection $ \Bc^{[i]} \cap \cpr $. 
\end{itemize}

\section{Completing the Attack for $ \lambda = 3 $}\label{secattaq}
 Following Coggia and Couvreur (\cite{coco}) for the $ 2 $-dimensional case, we can specify the goal of the attack as follows.\\
 \textbf{Objective:}\\
 \emph{To extract an alternative tuple $ \{ \gb_0', \gb_1 ', \gb_2 ', \gamma_1 ', \gamma_2 '  \} $ such that it satisfies:}
 \beq \label{eqgol}
 \cpr = \langle {\gb_0'}^{[i]} + \gamma_1 ' {\gb_1'}^{[i]} +   \gamma_2 ' {\gb_2'}^{[i]} \,\, \lvert \,\, i = 0,1, \cdots, n-k-1 \rangle 
 \eeq
 The analogous result was ingeniously achieved in the two-dimensional case by performing a semilinear transformation on the single parameter $ \gamma $ and setting up an equation to obtain the $ \gb $-parameters ($ ( \gb , \hb ) $ in \cite{coco}). We now show that the same trick works in the $ 3 $-dimensional case as well.
 \bprop\label{proptuple}
  Define $ \gamma_1 ', \gamma_2 ' \in \bfm \setminus \bF_q$ as follows:
  \beqn
  \gamma_{1} = \frac{a_{11}\gamma_{1} '  + a_{12}\gamma_{2 } ' + a_{13}}{a_{31}\gamma_{1} ' + a_{32}\gamma_{2} ' + a_{33}} ; \,\, \gamma_{2} = \frac{a_{21}\gamma_{1} ' + a_{22}\gamma_{2} ' + a_{23}}{a_{31}\gamma_{1} ' + a_{32}\gamma_{2} ' + a_{33}}.
    \eeqn
 where $ a_{ij} $ are the entries of a matrix $ \Ab \in GL_3 (\bF_q) $.\\
  Then the tuple $ \{ \gb_0', \gb_1 ', \gb_2 ', \gamma_1 ', \gamma_2 '  \} $ satisfies Equation (\ref{eqgol}) for the following choices:
 \beqn
\gb_0 ' = a_{33} \gb_0 + a_{13} \gb_1 + a_{23} \gb_2 ; \,\, \gb_1 ' = a_{31} \gb_0 + a_{11} \gb_1 + a_{21} \gb_2 ; \,\, \gb_2 ' = a_{32} \gb_0 + a_{12} \gb_1 + a_{22} \gb_2 .
 \eeqn
 \eprop
 \begin{proof}
 Substituting the values of $ \gamma_1, \gamma_2 $	in $ {\gb_0}^{[i]} + \gamma_1  {\gb_1}^{[i]} +   \gamma_2  {\gb_2}^{[i]} $ and rearranging in the form $ (\cdots)^{[i]} + \gamma_1 ' (\cdots)^{[i]} + \gamma_2 ' (\cdots)^{[i]} $ leads to the assertion. 
 \end{proof}
\qed
  It was further shown in \cite{coco} that the secret subspace parameter $ \gamma $ was the root of a polynomial with constituent factors of the form: $ X^{[i]} - X^{[j]} $. They established that performing a standard semilinear transformation on any root yielded another root - the projective linear group $ PGL(2, q) $ acts sharply transitively on the set of roots. In this section, we produce a bivariate polynomial of which each root pair is a tuple $ \{ \gamma_{1}, \gamma_{2} \} $ that can similarly lead to a valid secret key.
 
 \subsection{The Polynomial Equation for $ \{ \gamma_1, \gamma_2 \} $}\label{subsecuv}
 We now use the extraction subspaces to set up a polynomial equation for the tuple $ \{ \gamma_1, \gamma_2 \} $. Owing to the structure of the underlying Gabidulin codes, we can choose any element in $ \langle \gb_0 + \gamma_1 \gb_1 + \gamma_2 \gb_2  \rangle $ as a candidate for $ \gb_0 + \gamma_1 \gb_1 + \gamma_2 \gb_2  $. Moreover, for distinct integers $ i,j,k $ in the range $ [1 , n-k-1] $, we can show that (cf. Lemma $ 5 $ in \cite{coco}) $ \langle \gb_0, \gb_1, \gb_2 \rangle $ can be expressed as a direct sum as follows:
 \[ \langle \gb_{0} + \gamma_{1}^{[-i]} \gb_{1} +\gamma_{2}^{[-i]} \gb_{2} \rangle \oplus \langle \gb_{0} + \gamma_{1}^{[-j]} \gb_{1} +\gamma_{2}^{[-j]} \gb_{2} \rangle \oplus \langle \gb_{0} + \gamma_{1}^{[-k]} \gb_{1} +\gamma_{2}^{[-k]} \gb_{2} \rangle .\]
  This implies that, given a choice of $ (i,j,k) $  there exists a unique triple $ (\ub , \vb, \wb) $ such that: $ \ub + \vb + \wb = \gb_0 + \gamma_1 \gb_1 + \gamma_2 \gb_2 $, where $ \ub = k_1 (\gb_{0} + \gamma_{1}^{[-i]} \gb_{1} +\gamma_{2}^{[-i]} \gb_{2}) $, $ \vb = k_2 (\gb_{0} + \gamma_{1}^{[-j]} \gb_{1} +\gamma_{2}^{[-j]} \gb_{2}) $ and $ \wb = k_3 (\gb_{0} + \gamma_{1}^{[-k]} \gb_{1} +\gamma_{2}^{[-k]} \gb_{2}) $, for $ k_i \in \bfm $. Thus we have:
  \begin{equation}\label{eqcoeff}
  \begin{aligned}
  k_1 + k_2 + k_3 = 1; \\
  k_1\gamma_{1}^{[-i]} + k_2\gamma_{1}^{[-j]} + k_3\gamma_{1}^{[-k]} = \gamma_1; \\
  k_1\gamma_{2}^{[-i]} + k_2\gamma_{2}^{[-j]} +k_3\gamma_{2}^{[-k]} = \gamma_2. 
  \end{aligned}
   \end{equation} 

Solving the system of equations (\ref{eqcoeff}), we obtain:
\begin{equation}\label{eqki}
	\begin{aligned}
	k_1 = \frac{1}{\Delta} [(\gamma_1  \gamma_2^{[-j]} - \gamma_2  \gamma_1^{[-j]}) + (\gamma_2  \gamma_1^{[-k]} - \gamma_1  \gamma_2^{[-k]}) + (\gamma_1^{[-j]}  \gamma_2^{[-k]} - \gamma_1^{[-k]}  \gamma_2^{[-j]})];\\
	k_2 = \frac{1}{\Delta} [(\gamma_1  \gamma_2^{[-k]} - \gamma_2  \gamma_1^{[-k]}) + (\gamma_2  \gamma_1^{[-i]} - \gamma_1  \gamma_2^{[-i]}) + (\gamma_1^{[-k]}  \gamma_2^{[-i]} - \gamma_1^{[-i]}  \gamma_2^{[-k]})];\\
	k_3 = \frac{1}{\Delta} [(\gamma_1  \gamma_2^{[-i]} - \gamma_2  \gamma_1^{[-i]}) + (\gamma_2  \gamma_1^{[-j]} - \gamma_1  \gamma_2^{[-j]}) + (\gamma_1^{[-i]}  \gamma_2^{[-j]} - \gamma_1^{[-j]}  \gamma_2^{[-i]})];
	\end{aligned}
\end{equation} 
where \beq \label{eqdel}
 \Delta = (\gamma_1 ^{[-i]} \gamma_2^{[-j]} - \gamma_1^{[-j]} \gamma_2^{[-i]}) + (\gamma_1^{[-k]}\gamma_2^{[-i]}  - \gamma_1^{[-i]}  \gamma_2^{[-k]}) + (\gamma_1^{[-j]}  \gamma_2^{[-k]} - \gamma_1^{[-k]}  \gamma_2^{[-j]}).
 \eeq
 

\vspace{3.6 mm}
Denote the vector $ \ub $ obtained for the index set $ \{ i,j,k \}$ as $ \ub_{ijk} $. It is obvious that any pair $ (\ub_{ijk}, \ub_{ij'k'} )$ satisfies: $  \ub_{ijk} = \alpha \ub_{ij'k'}$, for some $ \alpha \in \bfm $, since $ \ub_{ijk}, \ub_{ij'k'} \in \langle (\gb_{0} + \gamma_{1}^{[-i]} \gb_{1} +\gamma_{2}^{[-i]} \gb_{2}) \rangle _{\bfm}$. As the coefficients $ k_i $ in (\ref{eqcoeff}) are dependent on $ i,j,k $ as well, we denote: 
\beqn
 \ub_{ijk} = k_{1}^{ijk} (\gb_{0} + \gamma_{1}^{[-i]} \gb_{1} +\gamma_{2}^{[-i]} \gb_{2}).
\eeqn
  This leads to an equation in terms of the coefficients used to describe the vector $ \ub $ as follows: $ k_{1}^{ijk} = \alpha k_{1}^{ij'k'} $. Substituting the values of $ k_1 $ for both sets of indices, using Equations (\ref{eqki}) and (\ref{eqdel}), we obtain a polynomial equation in $ \gamma_1, \gamma_2 $ over $ \bfm $.\\

As an illustration, the corresponding equation for $ \ub_{123} $ and $ \ub_{145} $ is :

\begin{multline} \label{eqset1}
\frac{1}{\Delta_{123}} [(\gamma_1  \gamma_{2}^{[-2]} - \gamma_2  \gamma_{1}^{[-2]}) + (\gamma_2  \gamma_{1}^{[-3]} - \gamma_1  \gamma_{2}^{[-3]}) + (\gamma_{1}^{[-2]}  \gamma_{2}^{[-3]} - \gamma_{1}^{[-3]}  \gamma_{2}^{[-2]})] = \\
\alpha \frac{1}{\Delta_{145}} [(\gamma_1  \gamma_2^{[-4]} - \gamma_2  \gamma_1^{[-4]}) + (\gamma_2  \gamma_1^{[-5]} - \gamma_1  \gamma_2^{[-5]}) + (\gamma_1^{[-4]}  \gamma_2^{[-5]} - \gamma_1^{[-5]}  \gamma_2^{[-4]}) ]
\end{multline}

where $ \Delta_{1jk} $'s are obtained by similar substitutions in (\ref{eqdel}) and $ \alpha \in \bfm^{\ast} $.\\
 
Raising both sides of (\ref{eqset1}) to $ q^5 $ and rearranging the factors, we have a polynomial equation in $ \gamma_1, \gamma_2 $. If we assign the pair of indeterminates $ X, Y $ to $ \gamma_1, \gamma_2 $ respectively, the resulting bi-variate equation has the form (for $ \alpha \in \bfm^{\ast} $):

\begin{multline} \label{eqxy}
[( X^ {[5]} Y^{[3]} - X^{[3]}Y^{[5]})
  + (X^{[3]}Y^{[2]} - X^{[2]} Y^{[3]}) + ( X^{[2]}Y^{[5]} - X^{[5]}Y^{[2]}) ] \\
 \times [( X^{[4]} Y^{[1]} - X^{[1]}Y^{[4]})  + ( X^{[1]}Y - XY^{[1]}) + (XY^{[4]} - X^{[4]}Y)] = \\
\alpha [ ( X^{[5]} Y^{[1]} - X^{[1]}Y^{[5]}) + ( X^{[1]}Y - XY^{[1]}) + ( X^{[5]}Y -X^{[5]}Y)]\\
\times [( X^{[4]} Y^{[3]} - X^{[3]}Y^{[4]}) + ( X^{[3]}Y^{[2]} - X^{[2]}Y^{[3]}) + (  X^{[2]}Y^{[4]} -X^{[4]}Y^{[2]}) ].
\end{multline}

To obtain alternative candidates for the original pair $ (\gamma_1, \gamma_2) $, therefore, we will examine the set of roots of a modified version of the above equation.

\subsection{Linear group action on the set of roots }\label{subsecinitpol}
Recall that in \cite{coco}, the parameter $ \gamma \in \bfm \setminus \bF_q $, which spans the secret subspace $ \Vc = \langle (1, \gamma) \rangle $, is a root of a polynomial $ P_{\gamma} (X)  = \frac{1}{(X^q - X)^{q+1}} Q_{\gamma} (X)
 $, where $ Q_{\gamma} (X) $ has terms which are products of the form: $ A(X) = X^{q^i} - X^{q^j}$. We term $ P_{\gamma} (X) $ as the \emph{reduced polynomial} for the case $ \lambda = 2 $, as this was obtained by removing all the linear factors over $ \bF_q $, with multiplicities, from $ Q_{\gamma} (X) $. As formulated in \cite{coco}, a valid alternative $ \gamma' $ is obtained by the following map: 
 \beq \label{eqsemlin}
 PGL(2,q) \times \bP(1,q) \rightarrow \bP(1,q); \,\,  \left(\begin{pmatrix}
 	a & c\\
 	b & d
 	\end{pmatrix}, [\gamma : 1] \right) \mapsto \left[ \frac{a\gamma +b}{c\gamma +d} : 1 \right]
 \eeq
 where $ ad -bc \neq 0 $. As shown in (\ref{eqsemlin}), this map can be interpreted as the action of the projective linear group $ PGL(2,q) $ on the projective space $ \bP(1,q) $, and any $ \gamma' = \frac{a\gamma +b}{c\gamma +d} $ may be chosen for $ \gamma $. Further, (cf. Lemma $ 6 $ and Proposition $ 5 $ in \cite{coco}) the transformation $ \gamma \mapsto \frac{a\gamma +b}{c\gamma +d} $ fixes the roots of $ A(X)=0 $. As $ \gamma \in \bfm \setminus \bF_q $, this implies that the set of roots of $ P_{\gamma} (X) = 0 $ is fixed as well. This leads to the conclusion that any root of $ P_{\gamma} (X) $ is a valid choice for the parameter $ \gamma $.\\
 It is evident that the above action is a collineation in two variables $ x_1,x_2 $, representing the general basis elements of a $ 2 $-dimensional $ \Vc $, which leads to a linear fractional transformation of the ratio $ \gamma = \frac{x_1}{x_2} $. In the $ 3 $-dimensional case, we consider a collineation in three variables $ x_1, x_2, x_3 $, leading to a linear fractional transformation on the $ 2 $ ratios: $ \gamma_1 = \frac{x_1}{x_3} $ and $ \gamma_2 = \frac{x_2}{x_3} $ given by:
 \beq \label{eq3semlin} 
 \gamma_{1}' = \frac{a_{11}\gamma_{1} + a_{12}\gamma_2 + a_{13}}{a_{31}\gamma_{1} + a_{32}\gamma_2 + a_{33}} ; \,\, \gamma_{2}' = \frac{a_{21}\gamma_{1} + a_{22}\gamma_2 + a_{23}}{a_{31}\gamma_{1} + a_{32}\gamma_2 + a_{33}}.
 \eeq
 
 Similar to the defining map in (\ref{eqsemlin}), the coefficients in (\ref{eq3semlin}) form a $ 3 \times 3 $ matrix $ \Ab = (a_{ij}) $ over $ \bF_q $, with non-zero determinant.\\
 One observes that the polynomial equation in (\ref{eqxy}), having $ ( \gamma_{1}, \gamma_2 ) $ as a root pair, is constituted of factors of the following form: 
 \beq \label{eqform}
 f^{(ijk)} (X,Y) :=
( X^ {[i]} Y^{[j]} - X^{[j]}Y^{[i]})
+ (X^{[k]}Y^{[i]} - X^{[i]} Y^{[k]}) + ( X^{[j]}Y^{[k]} - X^{[k]}Y^{[j]}) 
 \eeq
  It is required that the action induced by the `collineation matrix' $ \Ab $ should fix the set of roots of the polynomial of the form given in (\ref{eqxy}): a way to achieve that is to fix the set of roots of any term having the form given in (\ref{eqform}). To that end we have the following lemma.
  
\blem \label{lemselin}
Under the transformations 
\beqn  
X \mapsto \frac{a_{11}X + a_{12}Y + a_{13}}{a_{31}X + a_{32}Y + a_{33}} ; \,\, Y \mapsto \frac{a_{21}X + a_{22}Y + a_{23}}{a_{31}X + a_{32}Y + a_{33}}; \,\, a_{ij} \in \bF_q,
\eeqn
a polynomial $ f^{(ijk)} (X,Y) $ as given in Equation (\ref{eqform}) is transformed to:
\beq \label{eqtranspol}
\frac{\Delta_{\Ab}}{(a_{31}X + a_{32}Y + a_{33})^{[i] + [j] + [k]}}f^{(ijk) }(X,Y)
\eeq

where $ \Delta_{\Ab} $ is the determinant of $ \Ab = (a_{ij})$, a $ 3 \times 3 $ matrix over $ \bF_q $.

\elem
\begin{proof}
	Under the transformations, all the terms of the following forms have zero coefficient.
	\begin{enumerate}
		\item $ X^{[i]+[j]+[k]} $, $ Y^{[i]+[j]+[k]} $; 
		\item $ X^{[a]+[b]}Y^{[c]} $, $ Y^{[a]+[b]}X^{[c]} $, where $ (a,b,c) $ are permutations of $ (i,j,k) $;
		\item $ X^{[a]} $, $ Y^{[b]} $, where $ a $ and $ b $ run over $ i,j,k $.
	\end{enumerate} 
	The non-zero terms may be grouped as:\\
	 $ ( X^ {[i]} Y^{[j]} - X^{[j]}Y^{[i]})$, $(X^{[k]}Y^{[i]} - X^{[i]} Y^{[k]})$ and $( X^{[j]}Y^{[k]} - X^{[k]}Y^{[j]}) $,\\
	 each with coefficient: 
	 \beqn
	 \frac{\Delta_{\Ab}}{(a_{31}X + a_{32}Y + a_{33})^{[i] + [j] + [k]}}.
	 \eeqn
	 
\end{proof}
\qed

It is evident from (\ref{eqxy}) that the polynomial equation $ \Fc(X,Y) $ satisfied by $ ( \gamma_1, \gamma_2) $ is of the following form:
\beq\label{eqF}
\Fc(X,Y) = f^{(i_1 j_1 k_1) }(X,Y)f^{(i_2 j_2 k_2 ) }(X,Y) - \alpha f^{(i_1 j_2 k_2) }(X,Y)f^{(i_2 j_1 k_1 ) }(X,Y)
\eeq
where $ i_l > j_l > k_l ; \,\, l = 1,2$, with $ i_1 \geq i_2 $, the two sets of $ (i,j,k) $-indices being distinct, and $ \alpha \in \bfm^{\ast} $.\\
From Lemma \ref{lemselin}, it follows that, under the stated transformations,
\beq
\Fc(X,Y) \longmapsto \frac{\Delta_{A}^{2}}{(a_{31}X + a_{32}Y + a_{33})^{\Sigma}} \,\Fc(X,Y)
\eeq 
where $ \Sigma := [i_1] + [j_1] + [k_1] + [i_2] + [j_2] + [k_2]$.\\

For any pair $(\gamma_1, \gamma_2) $ such that $ \Vc = \langle 1, \gamma_1, \gamma_2 \rangle $ is a $ 3 $-dimensional vector space over $ \bF_q $, we have: $ (a_{31}\gamma_1 + a_{32}\gamma_2 + a_{33}) \neq 0$ when the $ a_{ij} $'s from $ \bF_q $ are not all zero. As any root pair of $ \Fc (X, Y ) $ sends $ [\Delta_{A}^{2}/(a_{31}X + a_{32}Y + a_{33})^{\Sigma}]\,\Fc(X,Y) $ to zero, the semilinear transformation on the two original basis elements $ \{\gamma_1, \gamma_2 \} $ produces another pair of basis elements which satisfies the same equation $ \Fc (X,Y) $. In the $ 2 $-dimensional case, the action of $ PGL(2,q) $ is sharply transitive on the points of $ PG(1,q) $. Hence one could conclude that all the roots of the reduced polynomial $ P_{\gamma} (X) $ for the single parameter $ \gamma $ belonged to the single orbit of the action. In the $ 3 $-dimensional case, we have established that the action  of $ PGL(3,q) $ on the points of $ PG(2,q) $ indeed maps one root of the \emph{initial  polynomial} $ \Fc (X,Y) $ to another. We next construct a counterpart of the reduced polynomial $ P_{\gamma} (X) $ that is analogously obtained from $ \Fc (X,Y) $.
\subsection{The reduced polynomial for $ \lambda = 3 $}
We first analyze the construction of the reduced polynomial $ P_{\gamma} (X) $ in the $ 2 $-dimensional case. To that end we have the following proposition.
\bprop \label{propgcd}
Denote the polynomial $ Q_{\gamma} (X) $ as $ Q_{\gamma} (X) = f_1(X) - \alpha^{q^3}f_2(X)$ where $ f_1(X) := (X^{q^3} - X^q) (X^{q^2} - X)$ and $f_2(X)  := (X^{q^3} - X) (X^{q^2} - X^q) $ and $ \alpha \in \bfm^{\ast} $. Then 
\beqn 
P_{\gamma} (X)  = \frac{Q_{\gamma} (X)}{\gcd (f_1(X),f_2(X))}
\eeqn

\eprop
\begin{proof}
As $ (X^{q^3} - X^q)  = (X^{q^2} - X)^q$, the roots of $ f_1 $ are the elements of $ \bF_{q^2} $, each counted with multiplicity $ q+1 $.\\
Further, $ (X^{q^3} - X) $ is the defining equation of $ \bF_{q^3} $ and $ (X^{q^2} - X^q) = (X^{q} - X)^q $. Thus the roots of $ f_2 $ are the elements of $ \bF_{q^3} $ and the elements of $ \bF_q $, the latter counted with multiplicity $ q $. As $ \bF_{q^3}\cap \bF_{q^2} = \bF_q $, it follows that 
\beq \gcd \label{2gcd} (f_1(X),f_2(X)) = (X^{q} - X)^{q+1}. \eeq
	
\end{proof}
\qed

We next attempt a similar reduction of the initial polynomial $ \Fc(X,Y) $, by first identifying the constituent polynomials as follows.
\beqn
\Fc(X,Y) = f_1(X,Y)f_2(X,Y) - \alpha f_3 (X,Y) f_4 (X,Y)
\eeqn 
where
\begin{multline}
f_1(X,Y) = [( X^ {[5]} Y^{[3]} - X^{[3]}Y^{[5]})
+ (X^{[3]}Y^{[2]} - X^{[2]} Y^{[3]}) + ( X^{[2]}Y^{[5]} - X^{[5]}Y^{[2]}) ] \\
f_2 (X,Y ) = [( X^{[4]} Y^{[1]} - X^{[1]}Y^{[4]})  + ( X^{[1]}Y - XY^{[1]}) + (XY^{[4]} - X^{[4]}Y)] \\
f_3(X,Y) = [ ( X^{[5]} Y^{[1]} - X^{[1]}Y^{[5]}) + ( X^{[1]}Y - XY^{[1]}) + ( X^{[5]}Y -X^{[5]}Y)]\\
f_4 (X,Y) = [( X^{[4]} Y^{[3]} - X^{[3]}Y^{[4]}) + ( X^{[3]}Y^{[2]} - X^{[2]}Y^{[3]}) + (  X^{[2]}Y^{[4]} -X^{[4]}Y^{[2]}) ].
\end{multline}
In order to reduce the initial polynomial, we will prove the following theorem on the existence of a common factor based on subsequent lemmas.
\bthm\label{thfacFc}
The initial polynomial $ \Fc(X,Y) $ has a factor of the following form:
\beq
\left[ \prod_{a \in \bF_q} (X + a) \prod_{b,c \in \bF_q} (bX + Y + c) \right]^{q^2 + 1}
\eeq
\ethm
The first lemma towards proving the theorem deals with the polynomials $ f_2(X,Y) $ and $ f_3 (X,Y) $.
\blem\label{f2f3}
The polynomials $ f_2(X,Y) $ and $ f_3 (X,Y) $ are divisible by 
\beqn f_0 (X, Y) = \prod_{a \in \bF_q} (X + a) \prod_{b,c \in \bF_q} (bX + Y + c) \eeqn
\elem
\begin{proof}

Examining the zeroes of the linear polynomials, it is readily established that every factor of the form $ X+a , \, a \in \bF_q$, or $ bX + Y + c, \, b,c \in \bF_q $, divides each of the polynomials: $ f_2(X,Y) $ and $ f_3(X,Y) $. Both the $ X $-degree and $ Y $-degree of $ f_0 (X,Y) $ equal $ q^2 $, while those for $ f_2 $ and $ f_3 $ are $ q^4 $ and $ q^5 $, respectively. The total degree of $ f_0 $ is $ q + q(q-1) + q = q^2 + q $, which is again less than $ q^4 + q $ for $ f_2 $ and $ q^5 + q $ for $ f_3 $. Hence the lemma.

\end{proof}
\qed

Next we examine the polynomial $ f_4 (X,Y) $ and prove that it divides $ f_5 (X,Y) $.
\blem\label{f4}
The polynomial $ f_4 (X,Y) = -[ f_0 (X,Y)]^{q^2}$ where $ f_0 (X,Y) $ is the product of linear factors as defined in Lemma \ref{f2f3}. 
\elem
\begin{proof}
Clearly $ \prod_{a \in \bF_q} (X + a) = X^q - X $.\\
We further have:
\beqn 
\begin{split}
\prod_{b,c \in \bF_q} (bX + Y + c) & = \prod_{b \in \bF_q} \prod_{c \in \bF_q}((bX + Y) + c)  \\ 
 & = \prod_{b \in \bF_q} ((bX + Y)^q - (bX + Y)) \\
 & = \prod_{b \in \bF_q} ( (Y^q - Y) + b(X^q -X))\\
 & = (X^q -X)^q \prod_{b \in \bF_q} ( Z + b), \,\, Z := \frac{Y^q - Y}{X^q -X}\\
 & = (X^q -X)^q (Z^q -Z)\\
 & = ((Y^q - Y)^q - (X^q -X)^{q-1} (Y^q - Y))\\
 & = ((Y^{q^2} - Y^q) - (X^q -X)^{q-1} (Y^q - Y)).
 \end{split}
 \eeqn
 Thus it follows that:
 \beqn 
 \begin{split}
 f_0 (X, Y) & = (X^q - X)((Y^{q^2} - Y^q) - (X^q -X)^{q-1} (Y^q - Y))\\
 & = ((Y^{q^2} - Y^q)(X^q - X) - (X^{q^2} -X^q) (Y^q - Y)). 
 \end{split} 
 \eeqn
 Therefore, raising to the $ q^2 $-th power, we obtain:
 \beqn 
 \begin{split}
 [f_0 (X, Y)]^{q^2} & = [(Y^{q^2} - Y^q)(X^q - X) - (X^{q^2} -X^q) (Y^q - Y)]^{q^2}\\
 & = [Y^{q^2}X^q - X^{q^2}Y^q + Y^{q}X - X^{q}Y + YX^{q^2} - Y^{q^2}X]^{q^2}\\
 & = -[X^{[4]} Y^{[3]} - X^{[3]}Y^{[4]} +  X^{[3]}Y^{[2]} - X^{[2]}Y^{[3]} +   X^{[2]}Y^{[4]} -X^{[4]}Y^{[2]} ]\\
 & = -f_4 (X,Y)\end{split}\eeqn

\end{proof}
\qed

\blem
The polynomial $ f_4 (X,Y) $ divides $ f_1 (X,Y) $; consequently $ f_1 $ contains all the factors of $ f_0 (X,Y) $ with multiplicity at least $ q^2 $.
\elem	
\begin{proof}
The polynomial $ f_1 (X,Y) $ can be rewritten in the following form:
\beqn
f_1 (X,Y)= - [(Y^{q^3} - Y^q)(X^q - X) - (X^{q^3} -X^q) (Y^q - Y)]^{q^2}
\eeqn
Clearly $ \prod_{a \in \bF_q} (X + a) = X^q - X $ is a factor of the polynomial within the brackets on the r.h.s.\\
It can be shown that any pair $ (x,y) \in \bF_q \times \bF_q $, which is a zero of any linear factor of the form $ bX + Y + c $, $ b,c \in \bF_q $, is a zero of the bracketed polynomial as well. Taking into account the degree of the bracketed polynomial, one can conclude that $ [f_0 (X,Y)]^{q^2}$ divides $ f_1 (X,Y) $.\\
The assertions then follow from Lemma \ref{f4} .
\end{proof}
\qed

\newpage

\textbf{Proof of Theorem \ref{thfacFc}:}\\
It follows from the preceding lemmas that both the terms $ f_1(X,Y)f_2(X,Y) $ and $ f_3(X,Y)f_4(X,Y) $ contain $ [f_0 (X,Y)]^{q^2 + 1} $ as a factor where
\beqn
f_0 (X, Y) = \prod_{a \in \bF_q} (X + a) \prod_{b,c \in \bF_q} (bX + Y + c)
\eeqn
This proves the theorem.\\{\flushright $ \square $\\}
In view of Theorem \ref{thfacFc}, we define the reduced polynomial $ P_r (X,Y) $ for our case as follows.
\begin{definition} 
	The reduced polynomial for $ \lambda =3 $ is defined by:
	\beq\label{eqred}
	P_r (X,Y) = \frac{\Fc(X,Y)}{\left[ \prod_{a \in \bF_q} (X + a) \prod_{b,c \in \bF_q} (bX + Y + c) \right]^{q^2 + 1}}
	\eeq
\end{definition}
	
Before moving on to the final steps of key-recovery for $\lambda =3$, we present a brief comparison of the reduced polynomials $ P_{\gamma} (X) $ and $ P_r (X,Y) $ in the form of a few observations as follows.
\begin{enumerate}
	\item The reduced polynomial $ P_{\gamma} (X) $, for the case $\lambda = 2 $, was obtained in \cite{coco} from the initial polynomial $ Q_{\gamma} (X) $ by dividing out the factor $ (X^{q} - X)^{q+1} $. This is equivalent to factoring out all distinct linear polynomials over $ \bF_q $, each with multiplicity $ q+1 $. The polynomial $ P_r (X,Y) $ is obtained as the result of an analogous reduction on the initial polynomial $ \Fc(X,Y) $ by factoring out all the distinct linear polynomials in $ X,Y $ over $ \bF_q $, each with multiplicity $ q^2 + 1 $ (cf. Theorem \ref{thfacFc}).
	\item The degree of $ P_{\gamma} (X) $ matches exactly with the cardinality of $ PGL(2,q) $. However, for $ \lambda = 3 $, the total degree of $ \Fc(X,Y) $ is $ q^5 + q^4 + q^3 + q $, which is reduced by $ (q^2 + q)(q^2 +1) = q^4 + q^3 + q^2 + q$ to yield $ q^5 - q^2 $. In this case, the total degree of the reduced polynomial does not equal but divides $ \lvert PGL(3,q) \rvert = q^8 -q^6 -q^5 + q^3  $.
	\item The reduction factor in the case $ \lambda = 2 $ was precisely the gcd of the two additive components of the initial polynomial $ Q_{\gamma} (X) $. In dealing with $ \Fc (X,Y) $, one possible way of reduction would have been to use a Gr\"{o}bner basis of the ideal generated by the additive components $ f_1 (X,Y)f_2 (X,Y) $ and $ f_3 (X,Y)f_4 (X,Y) $. Instead we have extracted a common factor and proceeded to reduce $ \Fc (X,Y) $ with it. But our simulations for small field sizes using Sage (\cite{Sage}) suggest that this factor is indeed the greatest common divisor over $ \bF_q $ of the additive components in the sense of polynomial factorization ( cf. for instance, \cite{vzGG}). So we have the following\\
	\textbf{Conjecture:} \\
	\emph{The polynomials $ f_1(X,Y)f_2(X,Y) $ and $ f_3(X,Y)f_4(X,Y) $ in $ \bF_q [X,Y] $ have a greatest common factor given by $ [ f_0 (X,Y)]^{q^2 + 1} $ }.
	
\end{enumerate}

\subsection{Completion of Key-recovery}
We begin outlining the final steps of the key-recovery by first examining the action of $ PGL(3,q) $ on the roots of the reduced polynomial $ P_r (X,Y) $ defined in (\ref{eqred}). In particular, we consider the action on the \emph{defining} root pair $ (\gamma_1, \gamma_2) $ such that $ \gamma_1, \gamma_2 \in \bfm \setminus \bF_q $, with $ \dim_{\bF_q} \langle 1, \gamma_1, \gamma_2 \rangle  = 3$, are used to define the initial polynomial $ \Fc (X,Y) $ (cf. Subsection \ref{subsecinitpol}).
\bprop \label{actred}
Let $ (\gamma_1, \gamma_2) $ be the defining root pair of $ P_r (X,Y) $. Then the following action of $ PGL(3,q) $  as defined in Lemma \ref{lemselin}, specified by the matrix $ \Ab = (a_{ij}) \in GL_3 (\bF_q) $, maps $ (\gamma_1, \gamma_2 ) $ to another root of $ P_r (X,Y) $:
\beqn  
\gamma_1  \mapsto \frac{a_{11}\gamma_1 + a_{12}\gamma_2 + a_{13}}{a_{31}\gamma_1 + a_{32}\gamma_2 + a_{33}} ; \,\, \gamma_2 \mapsto \frac{a_{21}\gamma_1 + a_{22}\gamma_2 + a_{23}}{a_{31}\gamma_1 + a_{32}\gamma_2 + a_{33}}. 
\eeqn 

\eprop
\begin{proof}
	Enough to prove that the pair $ (\gamma_1 ', \gamma_2 ') $, defined as follows, is a root.
	  \beqn  
	\gamma_1 ' = \frac{a_{11}\gamma_1 + a_{12}\gamma_2 + a_{13}}{a_{31}\gamma_1 + a_{32}\gamma_2 + a_{33}} ; \,\, \gamma_2' = \frac{a_{21}\gamma_1 + a_{22}\gamma_2 + a_{23}}{a_{31}\gamma_1 + a_{32}\gamma_2 + a_{33}}. 
	\eeqn
	We have: $ P_r (X,Y) = \Fc (X,Y) / G(X,Y)$, with (cf. Lemma \ref{f4})
	\beq\label{eqG}
	G(X,Y) = [f_0 (X, Y)]^{q^2 +1} = f(X,Y)f_4 (X,Y) 
	\eeq
	where
\begin{multline*}
	f(X,Y) = [X^{q^2}Y^q - X^q Y^{q^2} + X^{q}Y - XY^{q} + XY^{q^2} - X^{q^2}Y ];\\
	f_4 (X,Y) = [ X^{[4]} Y^{[3]} - X^{[3]}Y^{[4]} +  X^{[3]}Y^{[2]} - X^{[2]}Y^{[3]} +   X^{[2]}Y^{[4]} -X^{[4]}Y^{[2]}].	
\end{multline*}
If we perform the given transformations substituting $ X $ for $ \gamma_1 $ and $ Y $ for $ \gamma_2 $, then the application of Lemma \ref{lemselin} yields:
\beqn
\begin{split}
	f(X,Y) \longmapsto \frac{\Delta_{\Ab}}{(a_{31}X + a_{32}Y + a_{33})^{q^2 + q + 1}}f(X,Y);\\
	f_4 (X,Y) \longmapsto \frac{\Delta_{\Ab}}{(a_{31}X + a_{32}Y + a_{33})^{q^4 + q^3 + q^2}}f_4(X,Y);\\
	\Fc (X,Y) \longmapsto \frac{\Delta_{\Ab}^{2}}{(a_{31}X + a_{32}Y + a_{33})^{q^5 +q^4 + q^3 + q^2 + q +1}} \Fc (X,Y) 
\end{split}
\eeqn
Hence, under the transformations, 
\beq\label{eqproot}
\begin{split}
	P_r (X,Y) \longmapsto & \frac{1}{(a_{31}X + a_{32}Y + a_{33})^{q^5- q^2}} \frac{\Fc (X, Y)}{f(X,Y)f_4 (X,Y)}\\ & = \frac{1}{(a_{31}X + a_{32}Y + a_{33})^{q^5- q^2}} P_r (X,Y).
\end{split}
\eeq
Reverting to $ \gamma_1, \gamma_2 $ we, therefore, have:
\beqn
P_r (\gamma_1', \gamma_2') = \frac{1}{(a_{31}\gamma_1 + a_{32}\gamma_2 + a_{33})^{q^5- q^2}} P_r (\gamma_1,\gamma_2).
\eeqn
Given that $ (a_{31}\gamma_1 + a_{32}\gamma_2 + a_{33}) \neq 0 $ from the definition of $ \gamma_1, \gamma_2 $, the proposition follows. 
\end{proof}

The above result indicates that we could proceed with a root pair, say $ (\gamma_1', \gamma_2') $, of the polynomial $ P_r (X,Y) $ to extract the tuple $ \{ \gb_0', \gb_1 ', \gb_2 ', \gamma_1 ', \gamma_2 '  \} $ for key-recovery. For the sake of completeness, we briefly outline the key steps in the Coggia-Couvreur attack for $ \lambda =3 $ as follows.
\begin{enumerate}
	\item \emph{Relating $ \gb_0', \gb_1 ', \gb_2 ' $ to known parameters:} Let
	\beqn  
	\gamma_1  = \frac{b_{11}\gamma_1 ' + b_{12}\gamma_2 ' + b_{13}}{b_{31}\gamma_1' + b_{32}\gamma_2 ' + b_{33}} ; \,\, \gamma_2 = \frac{b_{21}\gamma_1' + b_{22}\gamma_2' + b_{23}}{b_{31}\gamma_1' + b_{32}\gamma_2' + b_{33}}. 
	\eeqn
	for some $ \Bb = (b_{ij}) \in GL_3 (\bF_q) $.\\ 
	 Then we can obtain expressions for $ \gb_0', \gb_1 ', \gb_2 ' $ in terms of $ \gb_0, \gb_1 , \gb_2  $ and $ \gamma_1 ', \gamma_2 ' $ in the manner of Proposition \ref{proptuple}, from the equation: 
	 \beqn 
	 \gb_{0}' + \gamma_{1}' \gb_{1}' +\gamma_{2}' \gb_{2}'  = \gb_{0} + \gamma_{1} \gb_{1} +\gamma_{2} \gb_{2}.
	 \eeqn 
	\item \emph{Solving for $ \gb_0', \gb_1 ', \gb_2 ' $:} The quantities $ \ub_{123}, \vb_{123} $ defined in Subsection \ref{subsecuv}, are used to compute $ (\gb_{0}' + \gamma_{1}'^{[-1]} \gb_{1}' +\gamma_{2}'^{[-1]} \gb_{2}') $ and $ (\gb_{0}' + \gamma_{1}'^{[-2]} \gb_{1}' +\gamma_{2}'^{[-2]} \gb_{2}') $ (vide Lemma \ref{lemselin}). Using, in addition, the known vector $ \gb_{0}' + \gamma_{1}' \gb_{1}' +\gamma_{2}' \gb_{2}' $, we can extract $ \gb_0', \gb_1 ', \gb_2 ' $.
	\item The previous steps outline the recovery of an alternate key in the form of the tuple $ \{ \gb_0', \gb_1 ', \gb_2 ', \gamma_1 ', \gamma_2 '  \} $. Using this alternate key in the formulation of Subsection \ref{subsecdefgi}, we can compute the dual $ \cpr $ in a similar manner as presented in \cite{coco}, and hence, decrypt the ciphertext.  
\end{enumerate}
\section{Conclusion}
We have extended the key-recovery attack on Loidreau's rank-metric scheme, which was proposed by Coggia and Couvreur and proven for dimension parameter $ \lambda = 2 $ , to cases with $ \lambda >2 $. Specifically, we have detailed the steps to identify the non-random structure (the so-called ``distinguisher'') of the dual of the public code, denoted $ \cpr $, for all values of $ \lambda \geq 3$. Further, we have extended the key-recovery attack to $ \lambda = 3 $.\\
This expands a successful attack on Loidreau's scheme when the underlying code rate is $ \geq 1 - \frac{1}{\lambda} $. It will be worthwhile to attempt a modification of the attack to work for lower rate codes as well, especially for increasing values of $ \lambda $. In another direction, Loidreau's rank-metric scheme claims resistance to Overbeck-type attacks, among a few other proposals. It is certainly of interest to revisit the formulation of ``distinguishers'' and key-recovery attacks on the other Overbeck-resistant rank-metric schemes in the light of this success against Loidreau's scheme.

\begin{acknowledgements}
	The author would like to thank Arnab Chakraborty and Mridul Nandi for several helpful discussions.
\end{acknowledgements}


\bibliography{bebina}

\end{document}